 \documentclass[final,5p,times,twocolumn]{elsarticle}

\usepackage{amssymb}
\usepackage{amsthm}
\usepackage{breqn}
\usepackage{hyperref}
\usepackage{bbm}
\usepackage{graphicx}
\graphicspath{ {./Figure/} }
\usepackage{tikz}
\usepackage{thmtools, thm-restate} 
\usepackage[utf8]{inputenc}
\usepackage[english]{babel}
\usepackage{comment}
\usepackage{cleveref}

\def\R{{\mathbb{R}}}   

\def\N{{\mathbb{N}}}
\def\E{{\mathbb{E}}}
\def\P{{\mathbb P}}
\def\1{{\mathbf{1}}}

\def\A{{\mathcal A}}  
\def\H{{\mathcal H}}   
\def\K{{\mathcal K}}   

\DeclareMathOperator*{\argmax}{arg\,max}

\declaretheorem[name=Definition,
refname={definition,definitions},
Refname={Definition,Definitions}]{defdef}
\declaretheorem[name=Remark,
refname={remark,remarks},
Refname={Remark,Remarks}]{defrmk}

\journal{Systems \& Control Letters}

\begin{document}

\begin{frontmatter}



\title{Stochastic Control for Organ Donations: A Review\tnoteref{t1}} 

\tnotetext[t1]{This material is based upon work supported by the National Science Foundation under Grant IIS-2123684 and the U.S. Air Force Office of Scientific Research under Grant FA95502010211.}


\author[inst1]{Xingyu Ren}

\affiliation[inst1]{organization={Department of Electrical and Computer Engineering \& Institute for System Research},
            addressline={University of Maryland, College Park}, 
            postcode={20742}, 
            state={MD},
            country={USA}}

\author[inst1,inst2]{Michael C. Fu}
\author[inst1]{Steven I. Marcus}

\affiliation[inst2]{organization={Robert H. Smith School of Business},
            addressline={University of Maryland, College Park}, 
            postcode={20742}, 
            state={MD},
            country={USA}}

\begin{abstract}
    We review the literature on individual patient organ acceptance decision making by presenting a Markov Decision Process (MDP) model to formulate the organ acceptance decision process as a stochastic control problem. Under the umbrella of the MDP framework, we classify and summarize the major research streams and contributions. In particular, we focus on control limit-type policies, which are shown to be optimal under certain conditions and easy to implement in practice. Finally, we briefly discuss open problems and directions for future research.
\end{abstract}




\end{frontmatter}


\section{Introduction}\label{intro}
Organ transplantation is a medical treatment where a transplant surgeon removes an organ from a donor's body and transplants it into the body of a recipient (or patient) whose corresponding organ is non-functioning, damaged or missing. Liver and kidney transplantations account for about $80\%$ of the total number of organ transplantations in the U.S. every year, as organ transplantation is the best option for end-stage kidney disease (ESKD) patients and the only therapy for end-stage liver disease (ESLD) patients. Organ transplantation is also applied in the treatment of end-stage lung disease, heart failure, diabetes (due to non-functioning pancreas) and other organ failures. Depending on the donor, a donated organ is classified as a deceased donor or living donor organ. Living donor organ transplantation  \footnote{In living donor organ transplantation, only part of a donor's organ is removed. For example, a living donor may be able to donate one of their kidneys, one liver lobe, a lung or part of the lung, part of the pancreas or part of the intestines. However, living heart donation is currently not feasible.}, as an alternative to deceased donor organ transplantation, has become popular in recent years due to the growing need for transplantation, the shortage of deceased donor organs and its satisfactory performance. The outcome of living donor liver transplantation is comparable to the outcome of deceased donor liver transplantation \cite{wan2014operative}. Living donor organs may even be preferable among ESKD patients, as living donor recipients in general have better quality of life and societal participation than deceased donor recipients \cite{de2013difference}. Though organ transplantation is the preferred option of ESLD and ESKD patients, risks include those associated directly with the transplant surgery itself, as well as acute and chronic rejection of the donor organ, infections and side effects of taking medications (for anti-rejection or immunosuppressants) \cite{danovitch2009handbook}.

Patients eligible for transplantation have to join the waitlist of the United Network for Organ Sharing (UNOS), which allocates donated organs across the U.S. Once an organ is available, eligible patients on the waitlist will be prioritized by a complex scoring and ranking system. An available organ will be offered to patients successively until it is accepted. It has to be transplanted into the recipient as soon as possible and will be discarded if no one accepts it by a certain deadline. The time during transport is called the ``cold ischemia time". The shorter the cold ischemia time is, the better the transplant outcome. Besides joining the UNOS waitlist, ESKD patients with incompatible directed \footnote{A directed donor specifies the patient, usually a parent, family member or friend, who will receive their donated organ. A non-directed living donor does not name or have an intended recipient.} living donors have the option of joining a kidney paired donation (KPD) program, where living donor kidneys are swapped such that each patient is paired with a compatible donor. 

Because there is a shortage of organ donors, patients often have to wait for a long time, which could be several months or even years, depending on the type of organ. According to the UNOS database \cite{optn2020}, newly listed patients for kidney transplantation have increased 21\% from 2010 to 2019, while liver listings increased 12\%, and heart and lung listings increased more than 30\%. The kidney waitlist is even six times longer than the number of transplantation performed in 2019, and waitlists for liver, heart and lung transplantations are at least double the number of corresponding transplantations performed. In addition, researchers observe that a large portion of kidney and liver offers are declined by at least one candidate before being accepted for transplantation, frequently due to unsatisfactory quality \cite{husain2019association, howard2002transplant}. This raises the question: how should an individual patient behave regarding the acceptance of an organ for transplantation?

Patients make their decision under various uncertainties, including their future health status, availability of organ offers, and outcomes of the transplant surgery. A good organ acceptance strategy should reduce patient's risk and improve their quality of life and total lifetime expectancy. For example, low quality organs are often rejected by relatively healthy patients, who are more willing to wait longer for a potentially better
offer than patients in poor health \cite{howard2002transplant}. However, patients will be more likely to accept poor quality organs if the probability of getting an offer decreases as the waitlist increases. A patient's personal preferences, including their risk preference and expectations for life and health, also plays an important role in decision making. From a policy maker's perspective, understanding patient behavior helps increase the efficiency of the organ allocation system and enhances the social welfare.

Markov Decision Processes (MDPs) \cite{bertsekas2020dynamic} provide a general modeling paradigm for dynamic decision making under risk and uncertainty, and various MDP models have been proposed to study individual patient liver and kidney acceptance \cite{alagoz2009optimizing}. These models can also be modified to study the acceptance of other organs, including pancreas, lung and heart. The primary stochastic elements to be modeled include the patient's medical status, availability of organ offers and outcomes of transplantation. The goal is then to determine an acceptance strategy that maximizes the benefit of an individual patient (the decision maker). Control limit-type policies have been shown to be practical and desirable, through both rigorous theoretical arguments and empirical studies. When discussing specific models, we will focus on the corresponding optimal policies and their underlying intuition. 

The rest of the paper is organized as follows: In \Autoref{mdpframe}, we present a general MDP framework to show how to formulate the individual patient organ acceptance as a stochastic control problem. In particular, we introduce the concept of a control limit policy. In \Autoref{models}, we present various specific MDP models under the umbrella of the MDP framework. We identify the primary stochastic elements of the organ acceptance problem from their models and discuss their corresponding optimal policies. Finally in \Autoref{future}, we point out potential open research problems and future directions.

\section{MDP framework}\label{mdpframe}
In this section, we first describe the individual patient organ acceptance problem and propose an MDP model to illustrate the basic idea of modeling the organ acceptance as a stochastic control problem. Then we introduce the concept of control limit policies discussed in the context of organ acceptance.
\subsection{Problem settings and model formulation}
Our study focuses on the individual patient decision making regarding the acceptance of an organ for transplantation. The decision problem is whether to accept the offered organ based on patient medical status and characteristics of the organ offer. If accepted, the patient will undergo the transplantation, and the decision process terminates, whereas if rejected, the offer will be discarded, and the patient will wait for the next offered organ. The future patient state, availability of organ offers and outcomes of transplantation are the primary stochastic elements to be modeled. The patient will make a decision based on both patient and organ states. The objective is to maximize the total benefit of patient, for instance, the total quality-adjusted life years (QALY), over the entire decision process.

Now we present an MDP framework to properly depict the above-mentioned ingredients. With the exception of one model, we consider discrete-time MDP models only, where the set of decision epochs is the natural numbers $\N$. At each epoch, both patient state and organ offer are updated, and the patient has to make a decision on acceptance.

\subsubsection*{State of patient and organ offer stochastic processes}

We define the following stochastic processes to describe the evolution of patient state and organ offer over time.
\begin{itemize}
    \item $\{h_t\}_{t\in \N}$: \textit{patient state}. $h_t\in S_H\subseteq\R_+$ is a scalar representing the medical status of patient. For example, $h_t$ could be a composite index computed by patient's lab test results and other characteristics. If the patient dies, we set $h_t=0$.
    \item $\{k_t\}_{t\in \N}$: \textit{organ offer state}. $k_t\in S_K\subseteq\R_+$ is a scalar representing the characteristics (e.g., the quality or match level) of the offered organ at time $t$. We assume that there is at most one organ offer arriving at each decision epoch. If there is no organ offer arriving at $t$, we set $k_t=0$.
\end{itemize}
The state space is $S_H\times S_K$.

\begin{defrmk}\label{offer}
The term ``offered organs" refer to organs offered to patients on the UNOS waitlists, predominantly deceased-donor organs. Note that at each epoch, patients may have multiple choices of organs even though we assume that at most one organ offer arrives. For example, patients with directed living donors (who are relatives of the patient in most cases) may simultaneously join the UNOS waitlist. They may choose either offered deceased-donor organs or those from directed living donors.
\end{defrmk}

\begin{defrmk}\label{scalarexp}
In practice, the UNOS and transplant surgeons use some (scalar) composite indices to measure the medical status of a patient or a donor organ. For example, they use the model for end-stage liver disease (MELD) score, which is a weighted average of some blood testing results, to measure how severe a patient’s liver disease is. Other examples include the estimated post transplant survival (EPTS) score, which measures the medical status of an ESKD patient, and the kidney donor profile index (KDPI) score, which evaluates the quality of a kidney offer.
\end{defrmk}

We assume that the patient has at most one directed living donor over the entire decision process, and the organ from the directed living donor is always available and has fixed characteristics $k_{LD}\in S_K$.

\begin{defrmk}\label{arrival}
For a continuous-time model, it is more natural to consider an increasing sequence of stochastic arrival times (e.g., arrival times of a Poisson process) of organ offers $\{U_n\}_{n\in \N}$. Organ offers are available only at the random instants $\{U_n\}_{n\in \N}$ at which the patient has to make a decision regarding acceptance. 

\end{defrmk}

\subsubsection*{Action \& Action space}
Define $\{a_t\}_{t\in\N}$ to be the patient \textit{actions} over time. For each $t\in\N$, $a_t\in\A=W \bigcup T$ where $\A$ is the \textit{action space} including
\begin{itemize}
    \item $W$: the set of \textit{non-transplantation} actions. For example, a ESKD patient may go on medication or dialysis;
    \item $T=\{T_D,T_{LD}\}$: the set of \textit{transplantation} actions where $T_D$ is to transplant with the organ offered by the UNOS and $T_{LD}$ is to transplant with the organ donated by the directed living donor (if there is one).
\end{itemize}

Since most models to be discussed in \Autoref{models} have finite state and action spaces, to simplify notation, we assume henceforth that the MDP model has finite state and action spaces, i.e., $S_H$, $S_K$ and $\A$ are finite sets. Extensions to more general settings are straightforward, in terms of modeling.
\subsubsection*{Dynamics}
At each decision epoch $t\in\N$, patient state $h_t$ evolves according to the Markov transition probability function $p(\cdot|\cdot,\cdot),~p: S_H\times S_H \times \mathcal{A}\mapsto [0,1]$, i.e., at $t+1$, patient state transitions to $h_{t+1}$ with probability (w.p.) $p(h_{t+1}|h_t,a_t)$. Denote by $\K(k_{t}|h_{t}),~\K:S_K\times S_H\mapsto [0,1]$, the (conditional) probability for a patient in state $h_{t}$ to receive an offer of state $k_{t}$. The distribution of organ offer state $k_{t}$ depends only on the current patient state $h_t$, not on the history of patient state and organ offer. $k_{t}$ is not \textit{directly} affected (controlled) by the patient's action (decision).

We set $\{(h,k)\in S_H\times S_K \ | \ h=0\}$ to be absorbing states, as $h=0$ represents death.

\subsubsection*{Rewards \& Objective functions}
At decision epoch $t$, a patient in state $h_t$ who takes a non-transplantation action $a_t\in W$ receives an intermediate reward $r(h_t,a_t)$, $r: S_H  \times W \mapsto \R_+$. The offered organ arriving at $t$ is no longer available to the patient after $t$. Set $r(0,a)=0,~\forall a\in W$, i.e., no reward accrues after the patient dies.

If the patient accepts an organ by taking $T_D$ or $T_{LD}$, the decision process terminates. When taking $T_D$, the patient receives the terminal transplantation reward $R_D(h_t,k_t)$, $R_D: S_H \times S_K \mapsto \R_+$; when taking $T_{LD}$, the patient receives the terminal transplantation reward $R_{LD}(h_t,k_{LD})$, $R_{LD}: S_H \times S_K \mapsto \R_+$, where $k_{LD}$, the characteristics of the organ from the directed living donor, is a constant. Reward functions $R_D$ and $R_{LD}$ take into account all the short-term and long-term effects of the transplantation, as the decision process terminates after the transplantation. Similarly, we set $R_D(0,k)=R_{LD}(0,k)=0,~\forall k\in S_K$.

This type of MDP model falls into a special class of MDP models called optimal stopping problems. 
Denote by $\tau=\min\{t\in \N \ | \ a_t\in T\}$ the random terminal time when the patient undergoes transplantation. Given a policy $\pi:\N\times S_H\times S_K\mapsto \A$, the expected total discounted reward at the start with initial state $(h_0,k_0)$ is given by
\begin{align}
\begin{split}\label{reward2}
    &~~~~f_\pi(h_0,k_0)\\= &\E\left( \sum_{t=0}^{\tau-1} \beta^tr(h_t,\pi(t,h_t,k_t))  + \beta^\tau R(h_\tau,k_\tau,\pi(\tau,h_\tau,k_\tau)) \ \Bigg| \ (h_0,k_0)\right),
\end{split}
\end{align}
where $\beta\in(0,1]$ is the discount factor and $R(h_t,k_t,a):=\1_{\{a=T_D\}}R_D(h_t,k_t) + \1_{\{a=T_{LD}\}} R_{LD}(h_t,k_{LD}) $ is the transplantation reward. 

A policy $\pi$ is called a \textit{stationary policy} if it does not depend on time, i.e., there exists a function $d:S_H\times S_K \mapsto \A$ such that $\pi(t,\cdot,\cdot)= d(\cdot,\cdot),~\forall t\in\N$. It can be shown that there exists a stationary optimal policy for the MDP model proposed in \Autoref{mdpframe} \cite{borkar1988convex}. Therefore, we will only consider {stationary policies} from now on. The goal is to find a policy $\pi^*\in\Pi$ such that
\begin{align*}
    \pi^* \in \argmax_{\pi\in\Pi} f_\pi(h_0,k_0),~\forall (h_0,k_0)\in S_H\times S_K,
\end{align*}
where $\Pi$ is the set of all the feasible stationary policies. Denote the maximum expected total discounted reward function (also known as the value function) by $V:=f_{\pi^*}$.




\subsection{Control limit policies}
Finding the optimal solution of this MDP problem translates into finding a partition of the state space $S_H\times S_K$, where each part of the partition is assigned an optimal action. For individual patient organ acceptance MDP models, many researchers have proposed optimal policies built on the idea of a \textit{control limit policy}.
\begin{defdef}\label{clpolicy}
Suppose that an MDP model has a one-dimensional state space $S\subset \R$ and an action space $\A$. A stationary policy $\pi:S\mapsto \A$ is called a \textit{control limit policy} if there exists a finite collection of intervals $\{I_i\}_{i=1}^n$ partitioning $\R$ and satisfying
\begin{enumerate}
    \item For any $a\in\A$, there exists at most one interval $I_i$ satisfying $\pi(s)=a,~\forall s\in I_i \bigcap S$;
    \item For any interval $I_i$, there exists $a\in \A$ such that $\pi(s)=a,~\forall s\in I_i\bigcap S$.
\end{enumerate}
Endpoints of intervals $\{I_i\}_{i=1}^n$ are called control limits.
\end{defdef}
The simplest and most common control limit policy, where $n=2$, takes the following form:
\begin{align}
\begin{split}\label{contrlimit}
    \pi(s)=\begin{cases}
    a_1 &\text{if } s<s^*,\\
    a_2 &\text{if } s\geq s^*.
    \end{cases}   
\end{split}
\end{align}
This control limit policy is easy to implement: if the value of state $s$ is less than control limit $s^*$, take action $a_1$; otherwise take action $a_2$. If there exists a control limit optimal policy, solving the MDP problem boils down to finding the optimal control limits, which can be solved efficiently. Control limit optimal policies exist in many classes of MDP problems, including machine maintenance and replacement, inventory control and asset allocation \cite{bertsekas2020dynamic}.

The MDP model we proposed has a vector-valued state $(h_t,k_t)$ to which \Autoref{clpolicy} doesn't apply. However, under suitable conditions, by fixing the value of one state variable and projecting the state space onto the other dimension, researchers have established optimal policies that take the form of a control limit policy (in one-dimension).
\begin{defdef}
A stationary policy $\pi: S_H\times S_K\mapsto \A$ is called a patient-based control limit policy if for each $k\in S_K$, there exists a finite collection of intervals $\{I^k_i\}_{i=1}^{n_k}$ partitioning $\R$ and satisfying
\begin{enumerate}
    \item For any $a\in\A$, there exists at most one interval $I^k_i$ satisfying $\pi(h,k)=a,~\forall h\in I^k_i \bigcap  S_H$;
    \item For any interval $I^k_i$, there exists $a\in \A$ such that $\pi(h,k)=a,~\forall h\in I^k_i \bigcap  S_H$.
\end{enumerate}
\end{defdef}

\begin{defdef}
A stationary policy $\pi: S_H\times S_K\mapsto \A$ is called an organ-based control limit policy if for each $h\in S_H$, there exists a finite collection of intervals $\{I^h_i\}_{i=1}^{n_h}$ partitioning $\R$ and satisfying
\begin{enumerate}
    \item For any $a\in\A$, there exists at most one interval $I^h_i$ satisfying $\pi(h,k)=a,~\forall k\in I^h_i \bigcap  S_K$;
    \item For any interval $I^h_i$, there exists $a\in \A$ such that $\pi(h,k)=a,~\forall k\in I^h_i \bigcap  S_K$.
\end{enumerate}
\end{defdef}

\begin{figure}[h]
    \centering
\includegraphics{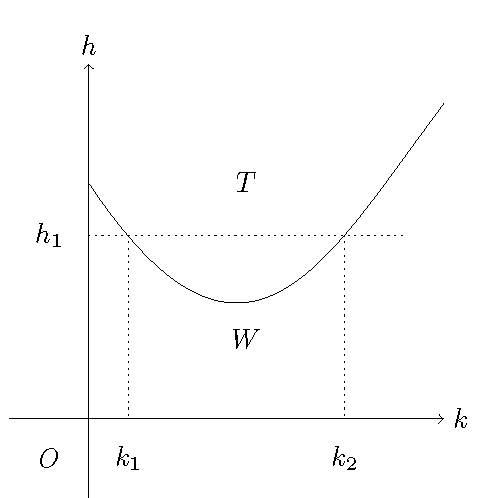}
\caption{Assume an optimal policy for an MDP model with $\A=\{W,T\}$ is depicted above. Then the corresponding patient-based optimal policy is a control limit policy, but the corresponding organ-based policy is not a control limit policy. For example, for fixed $h_1\in S_H$, action $W$ is specified in two disjoint intervals $[0,k_1]$ and $[k_2,\infty)$.}\label{counterex}
\end{figure}

Both control limit policies are intuitive: the decision maker should accept an organ offer if the offer is of sufficiently good quality, or the patient health is worse than some threshold. Note that existence of a patient-based control limit \textit{optimal} policy doesn't imply the existence of an organ-based control limit optimal policy, nor vice versa. A counterexample is provided in \Autoref{counterex}.

\begin{figure}[h]
    \centering
\includegraphics{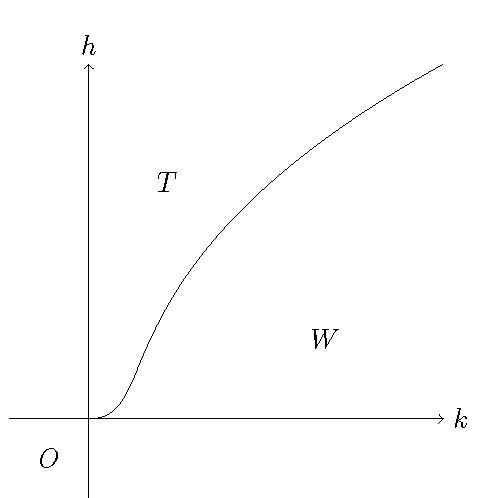}
\caption{For an MDP model with $\A=\{W,T\}$, if both patient-based and organ-based control limit optimal policies exist, then there exists an optimal policy such that states for which optimal actions are $W$ and $T$ are contained in two disjoint connected subsets.}\label{2acl}
\end{figure}

Action spaces of most MDP models to be reviewed consist of only two actions: reject the offer and wait $(W)$, and accept the offer for transplantation $(T)$. If both patient-based and organ-based control limit optimal policies exist, it is easy to show that there exists an optimal policy such that states for which optimal actions are $W$ and $T$ are respectively contained in two disjoint connected subsets of $\R^2_+$, as illustrated in \autoref{2acl}.

\begin{figure}[h]
    \centering
\includegraphics{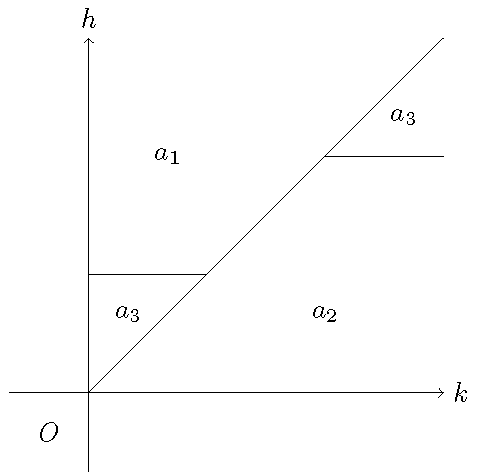}
\caption{For an MDP model with $|\A|= 3$, suppose that its unique optimal policy is depicted above. Both patient-based and organ-based control limit optimal policies exist. However, action $a_3$ is optimal in a disconnected region.}\label{3acl}
\end{figure}

Some researchers also establish a similar result when $|\A|=3$ \cite{alagoz2007choosing}. In general, if $|\A|\geq3$, the existence of both patient-based and organ-based control limit optimal policies does not guarantee that
there exists an optimal policy that allows $\R^2_+$ to be partitioned into at most $|\A|$ connected decision regions. A counterexample is provided in \Autoref{3acl}.

\section{Model review}\label{models}
In this section, we start with the only continuous-time model reviewed in this paper. Then we discuss a paradigm for discrete-time models in \Autoref{basicdis}, which is fairly close to the model proposed in \Autoref{mdpframe}. Discrete-time models in subsequent subsections can be viewed as variants of the paradigm.
\subsection{A continuous-time model}\label{contmodel}
\subsubsection*{Model formulation}
The continuous-time model in this section is the earliest one that formulates the kidney acceptance as an infinite-horizon optimal stopping problem \cite{david1985}. The patient's remaining {lifetime} at the beginning of the decision process is represented by a random variable $\tau$. The action space is $\A=\{W,T\}$, where $W$ is to reject the offer and wait, and $T$ is to accept the offer for transplantation. As mentioned in \Autoref{arrival}, the patient makes their decision only at {arrival times} of kidney offers $\{U_n\}_{n\in\N}$, as $W$ is the only available action at any $t\notin \{U_n\}_{n\in\N}$. States of kidney offers form a sequence of independent and identically distributed (i.i.d.), positive, bounded random variables $\{k_n\}_{n\in \N}$ having distribution function $F$. $\{k_n\}_{n\in \N}$, $\{U_n\}_{n\in\N}$ and $\tau$ are mutually independent. $\{(U_n,k_n)\}_{n\in\N}$ is a marked point process.

There is a terminal transplantation reward only. At time $U_j$, if the offer $k_{j}$ is accepted, the decision process terminates and the patient receives the terminal {reward} $\beta(U_j) k_j$, where $\beta:\R_+ \mapsto (0,1]$ is a nonincreasing discount function; otherwise, the offer is discarded, and the process continues until another offer arrives or the process terminates by itself (i.e., the patient's remaining lifetime is zero). Denote $1-\alpha_{j+1}=\P(\tau\leq U_{j+1} | \tau > U_j)$ the probability that the process terminates by itself before the $(j+1)^{th}$ arrival time $U_{j+1}$, given that the patient survives through $U_j$. 
\subsubsection*{Structure of the optimal policies}
It is proved that organ-based control limit optimal policies exist under three organ arrival patterns. Moreover, conditions that the optimal control limit has to satisfy are derived.
\begin{enumerate}
    \item \textit{Fixed arrival instants}. Suppose that $\{U_n\}_{n=0}^N$ is a finite sequence of constants for some $N\in\mathbb Z_+$ and the process terminates by itself at $U_N$ if the patient is still alive but no offer is accepted by $U_N$. By backward dynamic programming, $\lambda_N^j$, the maximum expected total discounted reward starting at $U_j$ if the $j^{th}$ offer is {rejected}, can be recursively defined by
    \begin{align}\label{david1}
        \lambda_N^j = \alpha_{j+1}\left( \lambda_N^{j+1}F\left(\frac{\lambda^{j+1}_N}{\beta_{j+1}}\right) + \int_{\frac{\lambda^{j+1}_N}{\beta_{j+1}}}^\infty \beta_{j+1}x dF(x) \right),
    \end{align}
    where $\beta_j:=\beta(U_j)$. Therefore, an optimal policy takes the following form: accept the $j^{th}$ offer if and only if $ k_j > \lambda^j_N/\beta_j$. Moreover, this result is extended to the corresponding infinite horizon problem by taking $N\rightarrow\infty$. Let $l_j=\lim_{N\rightarrow\infty} \lambda^j_N$ and $\gamma_j={l_j}{/\beta_j}$. Taking $N\rightarrow\infty$ in \eqref{david1},
    \begin{align*}
        l_j=\alpha_{j+1}\beta_{j+1} \left( \gamma_{j+1}F(\gamma_{j+1})+\int_{\gamma_{j+1}}^\infty x dF(x) \right).
    \end{align*}
    In this case, it is optimal to accept the $j^{th}$ offer if and only if $k_j > \gamma_j$.

    \item \textit{Deteriorating lifetime and renewal-type arrival of offers.} Suppose that $G$ is the distribution function of lifetime $\tau$ and interarrival times $\{(U_j-U_{j-1})\}_{j=1}^\infty$ are i.i.d. with distribution function $H$. If the process does not terminate and an offer of quality $k$ arrives at time t, $V(t,k)$, the maximum expected total discounted reward starting at $t$ is given by
    \begin{align}
        V(t,k)= \max(\beta(t)k,\lambda(t)),
    \end{align}
    where $\lambda(t)=\int_{0}^\infty \overline{G}(s|t) \left( \int_{0}^\infty V(t+s,y)dF(y) \right) dH(s)$ is the maximum expected total discounted reward when the offer is rejected. $\overline{G}(s|t)=\P(\tau>t+s|\tau>t)$ is the probability of survival beyond $t+s$, given that the process survives at $t$.
    
    If $G$ has the increasing failure rate (IFR) property, i.e., $\overline{G}(s|t)$ is nonincreasing in $t$ for any $s\geq 0$, there exists an optimal policy fully characterized by $\lambda(t)$: it is optimal to accept an offer of state $k$ at time $t$ if and only if $k>\lambda(t)/\beta(t)$. Moreover, $\lambda (t)$ is continuous and nonincreasing on $[0,\infty)$. $\lambda(t)$ monotonically decreasing is intuitive: as time passes, the probability that the process terminates by itself increases due to the IFR property of $G$. So, patients tend to tolerate and accept low quality organs, as they get zero reward if the process terminates by itself. IFR is a widely-used concept to describe the wearing out of a system; it is also a necessary condition for the optimality of the control limit policy to hold here.
    
    A special case studied is where lifetime $\tau$ has an Erlang distribution (which is IFR), the organ arrival process is a homogeneous Poisson process with intensity $\mu$, and discount function $\beta(t)\equiv 1$ \cite{bendersky2016deciding}. Each $k_n$ admits a finite number of values $\infty>x_1>\cdots>x_m> 0$ with probabilities $p_1,\cdots,p_m$ respectively. Since $\lambda(t)$ is monotonically decreasing, critical times when the acceptance criterion changes are defined by
    \begin{align*}
    t_i=\begin{cases}
    \max\{0,\lambda^{-1}(x_i)\} &\text{if } \inf_{t\in\R_+}\lambda(t)< x_i\\
    \infty &\text{otherwise}
    \end{cases},~i=1,\cdots,m.
    \end{align*}
    Critical times satisfy $0= t_1< t_2<\cdots < t_m\leq \infty$ where for each $i$, $t\in[t_i , t_{i+1}]$ implies $x_{i+1}\leq \lambda (t) \leq x_i$ (setting $x_{m+1}=0,~ t_{m+1}= \infty$). Therefore, there exists an optimal policy taking a simple form: During each time interval $[t_i, t_{i+1}),~i=1,\cdots,m$, accept the offer of value $x_k$ if and only if $k\leq i$. \Autoref{ctexample} provides an example of function $\lambda(t)$ and corresponding critical times. Another study discusses the related computational issues and provides an analytic expression for $\lambda(t)$ \cite{bendersky2016full}.
    
    \begin{figure}[h]
    \centering
\includegraphics{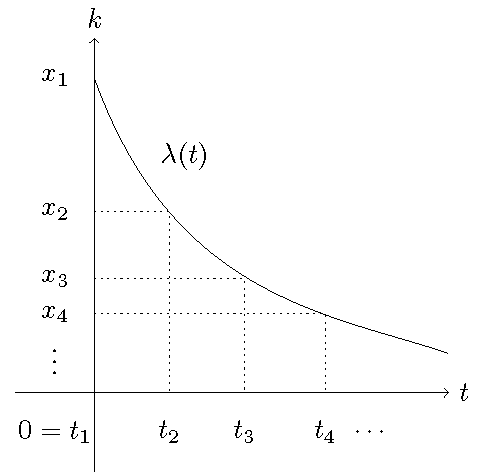}
\caption{Function $\lambda(t)$ and corresponding critical times. For example, during time interval $[t_2,t_3)$, the optimal strategy is to accept the organ offer if and only if it takes the value $x_1$ or $x_2$.}\label{ctexample}
\end{figure}
    
    \item \textit{Nonhomogeneous Poisson arrival process.} Suppose that the organ arrival process is a nonhomogeneous Poisson process with continuous intensity function $\mu(t)$. $\lambda(t)$, defined as in the previous case, satisfies an ordinary differential equation (ODE)
    \begin{align}\label{david2}
        \lambda'(t)=r(t)\lambda(t)-\beta(t)\mu(t)\int_{\frac{\lambda(t)}{\beta(t)}}^\infty \overline{F}(x) dx,
    \end{align}
    where $\overline F(x)=1-{F}(x)$, $r(t)={g(t)}/{(1-G(t))}$ is called the failure rate function and $g$ is the density function of $G$. Note that the previous case is a special case of this one if the organ arrival process is a homogeneous Poisson process with constant intensity $\mu$, and then $\lambda(t)$ can be obtained by solving \eqref{david2}.
\end{enumerate}


\subsection{Discrete-time model paradigm: Deceased donor liver} \label{basicdis}
The optimal acceptance of deceased donor livers for patients on the deceased donor waitlist is reviewed in this section. A discrete-time infinite-horizon MDP model is proposed \cite{alagoz2007determining}, in which the state is a two-dimensional vector describing both patient health and quality of the organ offer. The waitlist mechanism is implicitly modeled by the chance of a patient in various health states to receive organ offers of various qualities. Instead of including a terminal transplantation reward only as in \Autoref{contmodel}, the goal is to maximize the total quality adjusted lifetime expectancy (QALE) of the patient over the entire decision process. This model is also discussed thoroughly in another tutorial paper \cite{alagoz2010markov}.

At each time $t\in \N$, the state of the MDP is a random vector $(h_t,k_t)$, where $h_t\in S_H=\{1,\cdots,H,H+1\}$ represents the state of the patient. A larger value of $h_t$ implies worse patient health, and $H+1$ represents death. $k_t\in S_K=\{1,\cdots,K,K+1\}$ represents the state of the organ offer. A larger value of $k_t$ implies worse quality, and $K+1$ represents ``the organ is unavailable". Action space $\A=\{W,T\}$ is the same as \Autoref{contmodel}, where $T$ is to transplant, and $W$ is to reject and wait. At epoch $t$, if a patient in $h_t$ accepts the organ offer of state $k_t$, they receive the terminal transplantation reward $R(h_t,k_t)$, and the process terminates. Otherwise, that organ offer is no longer available after $t$, and the patient receives an intermediate reward $r(h_t)$ for being alive for one more epoch. At the next epoch $t+1$, the state vector transitions to $(h_{t+1},k_{t+1})$ w.p. $\H(h_{t+1}|h_t)\K(k_{t+1}|h_{t+1})$, where $\H(h_{t+1}|h_t)$ is the (conditional) probability that the patient is in $h_{t+1}$ at epoch $t+1$, given that they are in $h_t$ and do not undergo transplantation at epoch $t$; $\K(k_{t+1}|h_{t+1})$ is the (conditional) probability for a patient in state $h_{t+1}$ to receive an offer of state $k_{t+1}$ (recall that we assumed in \Autoref{mdpframe} that the distribution of organ offer $k_{t}$ depends only on the current patient state $h_t$). The decision process terminates if the patient undergoes transplantation or dies.

\begin{defrmk}
Note that the models in \Cref{basicdis,dismodel2,2donors} use the convention that larger values of $h_t$ (or $k_t$, respectively) implies worse patient health (or organ quality, respectively).
\end{defrmk}

$V(h,k)$, the maximum expected total discounted reward at the start with initial state $(h,k)$, satisfies
Bellman's equation
\begin{align}
\begin{split}
\begin{cases}
    V(h,k)=\max\{
    R(h,k),
    V^W(h,k)
    \} &\text{if }h\leq H\\
    V(H+1,k)=0
\end{cases}
    , ~\forall k \in S_K,
\end{split}
\end{align}
where \begin{align*}
    V^W(h,k)&=r(h)+\beta \sum_{h'\in S_H} V(h')  \H(h'|h),\\
     V(h)&=\sum_{k\in S_K} V(h,k)  \K(k|h).
\end{align*}
$V^W(h,k)$ is the maximum expected total discounted reward by choosing $W$ at the initial state $(h,k)$ and $V(h)$ is obtained by taking expectation of $V(h,k)$ over $k$.

Assuming the IFR property for $\H$ and monotonicity for reward functions, $V(h,k)$ is proved to be nonincreasing in both $h$ and $k$, i.e., the expected total discounted reward does not increase if the patient health deteriorates or the quality of organ offer drops. Moreover, both organ-based and patient-based control limit optimal policies exist under suitable conditions \cite{alagoz2007determining}. With $\A=\{W,T\}$, optimal policies are given by the following:
\begin{enumerate}
    \item Organ-based control limit optimal policy: for a patient in state $h$, it is optimal to accept the organ offer if and only if its quality is sufficiently good, i.e., if the organ state $k$ is less than some control limit $K^*(h),~K^*:S_H\mapsto S_K$.
    \item Patient-based control limit optimal policy: given an organ offer of state $k$, it is optimal to accept it if and only if the patient's health is sufficiently poor, i.e., if the patient state $h$ is greater than some control limit $H^*(k),~H^*:S_K\mapsto S_H$. 
\end{enumerate}
Both control limit policies are intuitive: the patient will accept the organ offer if either the organ is of sufficiently good quality, or the patient is in medically urgent condition (e.g., in bad health state). When both types of control limit optimal policies exist, it is straightforward to show that there exist $H^*$ and $K^*$ that are monotone and the inverse of each other. As a result, states for which optimal actions are transplant $(T)$ and wait $(W)$ are contained in two disjoint connected subsets of $\R^2_+$ (See \Autoref{2acl} for an example). 

\subsubsection*{Risk-sensitive patients}
This model can be modified to study risk-sensitive patients who want to maximize expected total utility rather than reward \cite{batun2018optimal}. An exponential utility function $u(x)=1-e^{-\gamma x},~\gamma>0$, is used to model risk-averse patients, where $u(x)$ is the patient's utility from surviving $x$ time epochs.

The state space and dynamics remain the same, but the reward now is measured in terms of patient lifetime instead of QALE. Denote the post-transplantation lifetime by a random variable $j$ which takes values in $S_T=\{0,1,\cdots,J\}$. For patients in state $h$, if they accept an offer of state $k$, a transplantation reward $j$ is conferred w.p. $J(j|h,k)$; otherwise, the transplantation does not happen and one unit reward accrues for being alive for one more epoch. The value function $V$ satisfies Bellman's equation:
\begin{align}
\begin{split}
\begin{cases}
    V(h,k)=\max \{V^T(h,k), V^W(h,k)\}  &\text{if } h\leq H\\
    V(H+1,k)=0
\end{cases} ,~\forall k\in  S_K,
\end{split}
\end{align}
where
\begin{align*}
    V^T(h,k)&=u^{-1}\left(\sum_{j\in S_T} J(j|h,k)u(j)\right), \\ V^W(h,k)&=u^{-1}\left(\sum_{h'\in S_H}\sum_{k'\in S_K} H(h'|h)\K(k'|h')u(1+V(h',k'))\right).
\end{align*}
$V^T(h,k)$ is the certainty equivalent \cite{howard1972risk} of the expected post-transplantation utility: if a patient in state $(h,k)$ chooses transplantation, their expected utility is $\sum_{j\in S_T} J(j|h,k)u(j)$, which is equal to the utility of being alive for $V^T(h,k)$ epochs. Similarly, $V^W(h,k)$ is the certainty equivalent of the expected utility from choosing $W$, and $V(h,k)$ is the certainty equivalent of the expected utility from making the optimal decision.

Monotonicity of value functions and existence of both types of control limit optimal policies can still be proved, but organ-based and patient-based results are slightly different \cite{batun2018optimal}:
\begin{enumerate}
    \item Organ-based results: for fixed $h$, $V(h,k)$ is convex and nonincreasing in $k$. Moreover, there exists an organ-based control limit optimal policy.
    \item Patient-based results: for fixed $k$, $V^T(h,k)$ is convex and nonincreasing in $h$. Moreover, there exists a patient-based control limit optimal policy.
\end{enumerate}

\subsection{Variant 1: Living donor liver}\label{dismodel2}
By removing the organ offer process $\{k_t\}_{t\in\N}$, the model in \Autoref{basicdis} can be modified to study the behavior of an ESLD patient with a directed living donor who wants to decide the optimal timing of transplantation \cite{alagoz2004optimal}. Assume that the living donor organ is always available and its quality is a constant. Dynamics and reward functions are similar to those in \Autoref{basicdis}: by taking $T$, a patient in $h_t$ receives a terminal transplantation reward $R(h_t)$ and the decision process terminates; otherwise, the patient who takes $W$ receives an intermediate reward $r(h_t)$, and at the next epoch, the patient state transitions to $h_{t+1}$ w.p. $\H(h_{t+1}|h_t)$. $V(h)$, the maximum expected total discounted reward at the start with initial state $h$, satisfies Bellman's equation \cite{nilim2005robust}
\begin{align}
\begin{split}
\begin{cases}
    V(h)=\max\{
    R(h),
    r(h)+\beta\sum_{h'\in S_H}V(h') \H(h'|h)
    \} &\text{if }h \leq H,\\
    V(H+1)=0   .
\end{cases}
\end{split}
\end{align}
Similar to \Autoref{basicdis}, it is proved that $V$ is nonincreasing and there exists a patient-based control limit optimal policy: it is optimal to accept the living donor organ if and only if the patient's health is sufficiently poor, i.e., the patient state $h$ is greater than some control limit $H^*$, which is a constant in this case.
\subsubsection*{A robust MDP formulation}
This model is further studied under uncertain patient state transition probabilities \cite{kaufman2017living}. A robust MDP model is proposed where the transition probability measure may vary within a set constructed by relative entropy (Kullback-Leibler divergence) upper bound. Specifically, Bellman's equation is given by
\begin{align}
\begin{split}
\begin{cases}
    V(h)=\max
    \begin{Bmatrix}
    R(h),\\
    r(h)+\beta \inf_{p\in \mathcal P(h)} \sum_{h'\in S_H}V(h') p(h')
    \end{Bmatrix}
     &\text{if } h\leq H\\
    V(H+1)=0
\end{cases},
\end{split}
\end{align}
where $V$ is the worst-case value function, $\mathcal P(h)$ is a set of probability measures over $S_H$, which describes transition probabilities of patient state starting from $h\in S_H$. $\mathcal P(h)$ is a ``ball" centered at the maximum likelihood estimate with radius to be the relative entropy upper bound in the space of probability measures:
\begin{align*}
    \mathcal{P}(h) = \{p\in \mathcal{M}(S_H) \ | \ D(p\|\hat{p}_h) \leq \beta(h))\},
\end{align*}
where $\mathcal{M}(S_H)$ is the set of all the probability measures over $S_H$, $D$ denotes the Kullback-Leibler divergence, $\hat{p}_h$ is the maximum likelihood estimate of transition probability measure starting from $h$, and $\beta(h)$ is called the level of ambiguity. The original model that directly uses the maximum likelihood estimate $\hat{p}_h$ as the true transition probability measure is called the myopic model. It can be shown that there exists a stationary optimal policy for the robust MDP model \cite{nilim2005robust}. The optimal policy of the robust model has the following properties \cite{kaufman2017living}:
\begin{enumerate}
    \item As the level of ambiguity increases, the optimal action is to transplant sooner (i.e., to transplant in healthier states).
    \item States in which the optimal action is transplantation under the myopic model is a subset of states in which the optimal action is transplantation under the robust model.
    \item If both robust and myopic models have (patient-based) control limit optimal policies, the control limit of the robust model is lower than or equal to the control limit of the myopic model. In other words, under the robust model, patients will take transplantation in healthier states, compared to the myopic model.
\end{enumerate}
\subsection{Variant 2: Choosing among living donor and deceased donor livers}\label{2donors}
\begin{figure}[h]
    \centering
\includegraphics{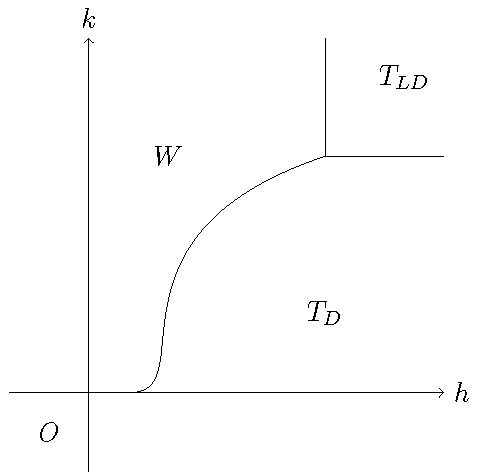}
\caption{An example of an AM3R policy that allows $\R_+^2$ plane to be partitioned into three connected decision regions.}\label{am3r}
\end{figure}

A patient with a directed living donor can simultaneously join the UNOS deceased donor liver waitlist \cite{alagoz2007choosing,alagoz2005choosing}, which is a combination of cases considered in \Autoref{basicdis} and \Autoref{dismodel2}. As in \Autoref{basicdis}, at each decision epoch $t\in \N$, the state of MDP is a random vector $(h_t,k_t)$ describing patient health and quality of the deceased donor organ offer. Assume that the organ from the directed living donor is always available and has fixed state $k_{LD}$. Now the action space is $\A=\{W,T_D,T_{LD}\}$, including waiting $(W)$, transplantation with the deceased donor organ $(T_D)$ and transplantation with the living donor organ $(T_{LD})$. The decision process terminates once the patient chooses $T_{LD}$ or $T_D$. If the patient accepts the deceased (or directed living) donor organ, a terminal transplantation reward $R_D(h_t,k_t)$ (or $R_{LD}(h_t,k_{LD})$) is conferred; otherwise, the patient receives an intermediate reward $r(h_n)$, and the state vector $(h_t,k_t)$ evolves in the same manner as in \Autoref{basicdis}. Now, Bellman's equation becomes
\begin{align}
\begin{split}
\begin{cases}
    V(h,k)=\max
    \begin{Bmatrix}
    R_D(h,k),\\
    R_{LD}(h,k_{LD}),\\
    r(h)+\beta \sum_{h'\in S_H}V(h') \H(h'|h)
    \end{Bmatrix}
    &\text{if } h\leq H\\
    V(H+1,k)=0
\end{cases},\\ \forall k \in S_K,
\end{split}
\end{align}
where $V(h)=\sum_{k\in S_K} V(h,k)  \K(k|h)$. The value function $V(h,k)$ is still nonincreasing in both $h$ and $k$. Since the patient has at most three options now, they show the existence of the following control limit-type optimal policies \cite{alagoz2007choosing}:
\begin{enumerate}
    \item At-most-2-region-organ-based (AM2RO) optimal policy : For fixed $h$, there exists a control limit $K^*(h),~K^*:S_H\mapsto S_K$ such that it is optimal to accept the deceased donor organ if and only if $k\leq K^*(h)$. Moreover, $a^*(h,K^*(h)+1)=a^*(h,K^*(h)+2)=\cdots=a^*(h,K+1)$, where $a^*(h,k)$ is the optimal action at state $(h,k)$.
    \item At-most-3-region optimal (AM3R) policy: For fixed $k$, there exists a control limit $H^*(k),~H^*:S_K\mapsto S_H$ such that it is optimal to wait if and only if $h\leq H^*(k)$. Moreover, there exists an AM2RO optimal policy. 
\end{enumerate}
Consequently, when an AM3R optimal policy exists, for each $a\in \A$, there exists a connected subset of $\R^2_+$ containing those states for which $a$ is optimal. Moreover, these subsets are mutually disjoint. An example is depicted in \Autoref{am3r}. The AM3R optimal policy has an intuitive explanation: for patients in good health, they should either wait or accept deceased donor organs of good quality; once their health status deteriorates below some threshold, they should transplant with the living donor organ unless the currently available deceased donor organ has better quality.

\subsection{Variant 3: Optimal timing to start dialysis treatment and accept a kidney offer}\label{dialysis}
In kidney transplantation, the timing to start dialysis has great impact on the outcome of transplantation \cite{churchill1997evidence,traynor2002early}. A discrete-time infinite horizon MDP model is developed to study the impact of dialysis \cite{fan2020optimal}. This model can still be regarded as a variant of the one in \Autoref{basicdis}, by incorporating actions, dynamics and rewards related to dialysis. Specifically, before undergoing transplantation or death, the patient could be on medication without starting dialysis or on dialysis. If the patient is on medication, three actions are available: medication $(M)$, dialysis $(D)$ and transplantation $(T)$. Once the patient starts dialysis, they have to continue on dialysis until transplantation or death. So, a patient on dialysis can only choose action $D$ or $T$. Patient state transition probability matrices and reward functions are different when being on medication or dialysis. Several kidney offer arrival patterns have been studied, and for each arrival pattern, criteria are proposed to decide the optimal action. However, those criteria are proved under restrictive conditions and are difficult to check in practice.



\begin{defrmk}
Settings of the continuous-time model in \Autoref{contmodel} \cite{david1985} are relatively simple. In particular, the patient state is represented by a single random variable representing remaining lifetime, which cannot model the progress of the disease, but enables the optimal control limit to be precisely described (e.g., by an ODE \eqref{david2}) and efficiently computed \cite{bendersky2016deciding}. Discrete-time models considering Markovian evolution of the patient's state establish the existence of control limit optimal policies but do not provide any guidance on computing an optimal policy. The basic model in \Autoref{basicdis} can be easily extended to study various settings in \Autoref{dismodel2} and \Autoref{2donors}, which also share similar structural properties, including the monotonicity of value functions and existence of control limit-type optimal policies.
\end{defrmk}

Now we present several studies which propose MDP models and use empirical or statistical methods to find policies that are reasonable and advantageous. These studies do not attempt to establish any theoretical optimality results.

\subsection{Variant 4: Continuous state space analog}
The discrete-time infinite horizon MDP model \cite{howard2002transplant} presented in this section can be viewed as a continuous state space analog of the one in \Autoref{basicdis}. At each decision epoch $t\in \N$, the state vector $(h_t,k_t)$ takes values in $(0,\overline{h}]\times(0,\overline{k}]$ where larger values of $h_t$ (or $k_t$, respectively) implies better patient health (or organ quality, respectively). Assume that $\{k_t\}_{t\in \N}$ is an i.i.d. sequence, also independent of $\{h_t\}_{t\in \N}$, with distribution function $G$. If a patient rejects an organ offer, their state $h$ evolves as a Markov process specified by transition density function $f(\cdot|h),~f:(0,\overline{h}]\times(0,\overline{h}]\mapsto \R_+$. If the patient accepts an organ offer of value $k$, w.p. $p(h,k)$, the transplantation is a success that confers a terminal reward of $B$ to the patient at the next epoch; otherwise, the failed transplantation results in immediate death with zero reward. The patient's single period reward for being alive is $u$. The process terminates when the patient undergoes transplantation or dies. So, starting at state $(h,k)$, the expected total reward from accepting an organ is
\begin{align*}
    V^{T}(h,k)=p(h,k) B,
\end{align*}
and the expected total reward from rejecting an organ and continuing to wait is
\begin{align*}
    V^W(h,k)=\int_0^{\overline{k}} \int_0^{\overline{h}} V(h',k')f(h'|h) dh' dG(k'),
\end{align*}
where $V(h,k)$ satisfies Bellman's equation
\begin{align*}
    V(h,k)=u+\beta \max(V^{T}(h,k),V^W(h,k)).
\end{align*}
An organ-based control limit policy is proposed: there exists an organ quality control limit function $k^c:(0,\overline{h}]\mapsto (0,\overline{k}]$ satisfying $V^{T}(h,k^c(h))=V^W(h,k^c(h)),~\forall h$, such that the organ should be accepted if $k>k^c(h)$. An empirical study shows that the control limit function $k^c(h)$ decreases when the patient state $h$ decreases (i.e., patient health deteriorates) or the waitlist increases. Their results coincide with and explain the behavior of patients in practice very well:
\begin{enumerate}
    \item Low quality organs are rejected by relatively healthier patients.
    \item As the waitlist increases, the use of poor quality organs is increased, since the chance for a patient to receive organ offers decreases.
\end{enumerate}

\subsection{Other variants}
Two studies presented in this section have state spaces entirely different from \Autoref{mdpframe}. The first study \cite{ahn1996involving} proposes a discrete-time infinite horizon MDP model where the patient state space $S_H$ consists of five possible stages in the kidney transplantation procedure: alive on dialysis waiting for transplantation $(S_1)$, not eligible for transplantation $(S_2)$, received a functioning transplant $(S_3)$, transplant failed $(S_4)$, and death $(S_5)$. Organ offer states $\{k_n\}_{n\in\N}$ are i.i.d., also independent of patient state. Action space is $\A=\{W,T\}$, where $W$ is to wait, and $T$ is to transplant, which is available only in $S_1$. In states other than $S_1$, $W$ is the only option. 

An organ-based control limit-type policy based on the one-year graft survival rate is proposed: the patient should accept an organ offer if its one-year graft survival rate exceeds the prescribed threshold. The one-year graft survival rate and its minimum acceptable criterion can be estimated from characteristics of both the patient and offered organ by empirical formulas. It is notable that this estimation takes into consideration not only the patient's medical status but also their personal preference.

Once all the model parameters are specified, the patient's QALE in each state can be estimated by stochastic simulation. Simulation results show that patients who are in good medical status or expect relatively high quality of post-transplant life can be selective on which organs to accept for transplantation.

Finally, we describe a study on deciding the order of deceased donor and living donor kidney transplantation for a pediatric patient with a directed living donor \cite{van2015choosing}. Similar to the other model, the MDP considers five stages of the treatment procedure: on the waitlist, post-transplantation with a deceased donor kidney, post-transplantation with a living donor kidney, after two graft failures, and death. The arrival and acceptance of organ offers are implicitly modeled by transition of the state. Two strategies are studied: living-donor-first followed if necessary by deceased donor retransplantation versus deceased-donor-first followed if necessary by living donor (if still available to donate) or deceased donor (if not) retransplantation. Once the patient chooses one strategy, the transition probability can be estimated by empirical formulas. Patient survival for the subsequent 20 years (which can be viewed as the patient's total reward) is predicted by stochastic simulation. Simulation results indicate that it is better for highly sensitized patients to choose deceased-donor-first strategy. For other patients, living-donor-first strategy is preferred. 

\section{Future research}\label{future}
We conclude our review by suggesting several future research questions and directions.

\textit{Compute the optimal control limit policy}. Since closed-form expressions for control limit optimal policies (like \eqref{david2}) are rare, numerical solutions by simulation-based methods may be required, where multiple sample paths are generated by stochastic simulation to estimate the performance of a given policy.

\textit{Consider more detailed patient and organ states.} States of most MDP models we discussed are two-dimensional vectors that describe patient and organ characteristics. Such oversimplified settings make it easier for researchers to derive structural results, including control limit-type policies. The real medical decision-making procedure is more complicated. Instead of a simple composite index, a vector of detailed features of the patient and organ donor including their ages, blood types and lab test results may be a more appropriate state definition for the MDP model. Structural results for such detailed models are likely to be more challenging to find and prove.

\textit{Consider multiple treatment options}. Most currently available MDP models have the following restrictions:
    \begin{itemize}
        \item The decision process terminates after a transplantation happens, and all the post-transplantation effect is summarized by a terminal reward. These decision models are less helpful for patients who are likely to experience retransplantations. For example, in \Autoref{2donors}, patients with directed living donors also enroll in the UNOS deceased donor waitlist. They may choose living-donor-first followed if necessary by deceased donor retransplantation, or the other way around. A model that considers explicitly retransplantation might be more appropriate in this case.
        \item In most MDP models, there are only two options available (wait or transplant). Currently existing studies considering multiple treatment options are not yet satisfactory. For example, results in \Autoref{dialysis} are derived under restrictive assumptions and hard to check in practice; the model in \Autoref{2donors} assumes that the living donor organ is always available with static quality, while in practice, the availability of a living donor is uncertain as time passes. How to properly model multiple treatment options and derive corresponding structural results could be a future research problem.
    \end{itemize}
    
\textit{Distinguish different organs}. Though most models are built in the context of kidney or liver transplantation, they don't stress features of different types of organs. As pointed out earlier, states of most MDP models are two-dimensional vectors representing patient and organ states, which are applicable in modeling any type of organ acceptance. Future research could aim to capture more specific characteristics of each type of organ transplantation. For example:
    \begin{itemize}
        \item Human leukocyte antigen (HLA) incompatibility is a major barrier in kidney and heart transplantation, but the role of HLA is less important or unclear in other types of organ transplantation. To receive an incompatible kidney, ESKD patients have to undergo desensitization treatment \cite{marfo2011desensitization}. Some recent work in \cite{ren2022optimal} studies the optimal acceptance of (possibly) incompatible kidneys by explicitly incorporating the incompatibility as a state variable and establishes control limit optimal policies.
        \item Hearts must be donated by recently deceased donors, i.e., there is currently no way to utilize living heart donors, whereas living donors can be utilized for all other organs. For instance, most people can get by with only one lung or kidney, which allows them to donate the other; for living liver donors, they only donate a portion of their livers, which will recover within several weeks after the transplant surgery because of remarkable regenerative capacities of human livers. 
        \item ESKD patients may choose a lifetime on dialysis, as an alternative to kidney transplantation. For other end-stage diseases, organ transplantation is considered necessary when the disease cannot be controlled by other treatment options.
    \end{itemize}
    
\textit{Consider multi-organ transplantation}. Combined pancreas-kidney and liver-kidney transplantation are major types of multi-organ transplantation \cite{optn2019}. Pancreas transplantation is a choice for people with type 1 diabetes. If they have kidney failure from diabetes, they may also accept a kidney transplant at the same time. Other types of multi-organ transplantation, including combined heart-lung, heart-kidney, lung-liver-pancreas, liver-pancreas-kidney and heart-lung-liver transplantation, are also performed in practice. Such settings may require new modeling, analysis and solutions.

\textit{Include patient personal preference.} For example, \Autoref{basicdis} considers risk-averse patients by applying an exponential utility function and then studies a risk-sensitive MDP \cite{batun2018optimal}. Such methods can also be applied to study risk-tolerant patients. Future research could also study other risk measures, as well as other types of preference incorporating financial considerations and quality of life.

\textit{Design organ allocation policies.} This review focuses on the individual patient behavior, with the objective of maximizing individual patient benefit, whereas a system view would consider the design of organ allocation policies \cite{roth2007efficient,kessler2012organ} where multiple patients are included and the decision-making process of individual patients could be modeled via game theory, e.g., using Markov game models.

 \bibliographystyle{elsarticle-num} 
 \bibliography{cas-refs}





\end{document}